\documentclass{article}
\usepackage{graphicx} % Required for inserting images
\usepackage{amssymb}
\usepackage{amsmath}
\usepackage{comment}
\usepackage{caption}
\usepackage[numbers]{natbib}
\usepackage{xcolor}
\usepackage{arxiv}
\newtheorem{theorem}{Theorem}
\newtheorem{proof}{Proof}

\usepackage{url}
\usepackage[colorinlistoftodos,textsize=tiny]{todonotes}
\newcommand{\Comments}{1}
\newcommand{\mynote}[2]{\ifnum\Comments=1\textcolor{#1}{#2}\fi}
\newcommand{\mytodo}[2]{\ifnum\Comments=1%
	\todo[linecolor=#1!80!black,backgroundcolor=#1,bordercolor=#1!80!black]{#2}\fi}

\ifnum\Comments=1               % fix margins for to-do notes
\fi

\DeclareMathOperator{\atan2}{atan2}

\newtheorem{remark}{Remark}%

\begin{document}
\title{On  the application of 
Jammalamadaka-Jimenez Gamero-Meintanis test for circular regression model assessment}

\fancyhf{} % Clear all header and footer fields
\fancyhead[L]{On  the application of 
JJGM test for circular regression model assessment} % Left header with the short title
\fancyhead[R]{A Preprint} 

%[On  the application of 
%JJGM test for circular regression model assessment]
\author{K. Halaj, B. Klar, B. Milo\v sevi\' c, M. Veljovi\' c}
\date{}
\maketitle

\begin{abstract}
We study a circular-circular multiplicative regression model, characterized by
an angular error distribution assumed to be wrapped Cauchy. We propose a
specification procedure for this model, focusing on adapting a recently proposed
goodness-of-fit test for circular distributions. We derive its limiting properties and
study the power performance of the test through extensive simulations, including
the adaptation of some other well-known goodness-of-fit tests for this type of
data. To emphasize the practical relevance of our methodology, we apply it to
several small real-world datasets and wind direction measurements in the Black
Forest region of southwestern Germany, demonstrating the power and versatility
of the presented approach.
\end{abstract}

\section{Introduction}\label{sec1}

Circular data has applications in many fields. For example,  meteorologists use them to analyze phenomena such as wind direction and ocean currents, ecologists study the direction of animal movement, and economists study calendar effects. 
However, due to the difference in topology from real-line data, a special methodology is required for such data. A detailed review can be found in \cite{pewsey2021recent}, including what is done for regression models. Here we focus on the circular-circular regression proposed by \cite{kato2008circular}, defined as follows.

Let $Y$ be a random variable with the density 
\begin{align}\label{wcdensity}f(y,{\delta}) = \frac{1}{2\pi}\frac{|1-|\delta|^2|}{|y-\delta|^2},\; y \in \Omega=\{z \in \mathbb{C}: |z|=1\},\; |\delta|\neq 1,\end{align}
where $\mathbb{C}$ is the set of complex numbers and $\delta=e^{i\mu-\gamma}$, for $\mu\in \mathbb{R}$ and $\gamma>0$.
Then we say that $Y$ has the wrapped Cauchy distribution (with mean direction equal to $\mu$) and write $Y \sim WC(\mu,\delta)$. If $\mu = 0$, $\delta$ is in $(0,1)$. If $\delta = 0$, $Y$ has a uniform distribution on the circle, and if $\delta$ tends to one, it will be concentrated in the mean direction.

\begin{comment}
where $\mathbb{C}$ is the set of complex numbers.
Then we say that $Y$ has the wrapped Cauchy distribution (with mean direction equal to zero) and write $Y \sim WC(\delta)$. If $\delta = 0$, $Y$ has a uniform distribution on the circle, and as $|\delta|$ tends to one, it becomes concentrated at the mean direction.
\end{comment}
\begin{remark}
The random variable $Y$ can be seen as a circular variable since it is a complex number with modulus 1, and is therefore completely defined by the angle in polar representation.
\end{remark}

In \cite{kato2008circular} the following regression model with wrapped Cauchy errors, was proposed.
Let $(x_1,Y_1),\ldots,(x_n, Y_n)$ be a sequence of independent circular observations such that
%The regression model proposed in \cite{kato2008circular}, is defined by
\begin{equation}\label{model}
    Y_i = \beta_0 \frac{x_i+\beta_1}{1+\overline{\beta_1}x_i}\varepsilon_i,\quad x \in \Omega, \;\varepsilon_i \sim WC(\delta), \;\text{ for some } 0 < \delta < 1,
\end{equation}
where $\beta_0 \in \Omega$, $\beta_1 \in \mathbb{C}$ and $WC(\delta)$ is short for the wrapped Cauchy $WC(0,\delta)$. Here, $\beta_0$ is a rotation parameter; for the interpretation of $\beta_1$, see \citet[p. 635]{kato2008circular}. 
$\varepsilon_1,\ldots,\varepsilon_n$ are also assumed to be independent.

The main goal of this paper is to provide a powerful diagnostic tool for testing the following assumption about the error distribution:
\begin{align}\label{hipoteza}
    H_0: \varepsilon_i\sim WC(\delta), \text{ for some } \delta\in (0,1),
\end{align}
against the alternative that the errors do not follow a $WC(\delta)$ distribution for any $\delta\in (0,1)$.

Since the errors $\{\varepsilon_i\}$ are not observable, a natural approach, though not yet used in the context of circular regression models, is to apply the test originally designed for i.i.d. data to model residuals in either complex form or their angular counterparts given by
\begin{align}
   \label{reziduali1} \widehat{\varepsilon}_j&=\frac{Y_j}{\widehat{Y}_j},\\
    \label{reziduali2}\widehat{\theta}_{\varepsilon_j}&=\arg(\widehat{\varepsilon}_j),
    %\widehat{\theta}_{\varepsilon_j}&=(\arg(\widehat{\varepsilon}_j)+2\pi)mod 2\pi
\end{align}
respectively, and $\hat{Y}_j$ being the regression estimate (see (\ref{hat-Y}) below). Defining residuals in this way reflects the multiplicative nature of the model. This approach is known as the gold standard for various regression models for linear data types (see, e.g., \cite{gonzalez2013updated} and \cite{meintanis2014class}). Here we apply goodness-of-fit tests with arbitrary circular distributions proposed in \cite{rao2019class} and explore their properties in this setting. The reason for focusing on this statistic, among many proposed for circular data, lies in its dominance over many alternative distributions.

The paper is organized as follows. In Section \ref{sec: teststat} we present the adaptation of the test statistic from \cite{rao2019class}. Its asymptotic properties are discussed in Section \ref{sec: asymptotic}. The proof of the main result is postponed to the Appendix \ref{proof}.
The powers of the test are compared with those of several competitors in Section \ref{sec: powers}. Section \ref{sec: realdata} contains several real data examples with small sample sizes and an application to wind direction measurements that clearly illustrate the tests' applicability.

\section{Test statistic}\label{sec: teststat}
As mentioned above, the main idea for testing the hypothesis \eqref{hipoteza} is to apply tests on the model residuals defined in \eqref{reziduali1} or \eqref{reziduali2}. 
Although other estimators could be applied, we follow the original approach of \cite{kato2008circular} and proceed with the maximum likelihood estimator (MLE) by considering all quantities in \eqref{model} in angular form.
Let  $(x_j,Y_j) = (e^{i\theta_{x_j}}, e^{i\theta_{Y_j}})$ and $(\beta_0, \beta_1 ) = (e^{i\theta_0}, r  e^{i\theta_1})$, where $r>0, 0 \leq \theta_0, \theta_1 \leq 2\pi$. Then
the unknown parameters $v =(\theta_0,\theta_1,r,\delta)$ are obtained by maximizing the log-likelihood function 
\begin{align*}
    \log L(v)&=n\log(1-\delta^2)\\&-\sum_{j=1}^n\log(1-2\delta\cos(\theta_{y_j}-\theta_0-\theta_{x_j}+2\arg(1+r e^{i(\theta_{x_j}-\theta_1)}))+\delta^2)+const.
\end{align*}
The estimates are denoted by $\widehat{v} =(\widehat{\theta}_{0},\widehat{\theta}_{1},\widehat{r},\widehat{\delta})$. The estimated regression function at $x=x_j$ is then equal to
\begin{align} \label{hat-Y}
    \hat{Y}_j=\hat{\beta}_0 \frac{x_j+\hat{\beta}_1}{1+\overline{\hat{\beta}_1}x_j}
\end{align}
and residuals can be calculated. 

The characteristic function of the wrapped Cauchy distribution is given by 
$\varphi(t,\delta)= \mathbb{E}[e^{it\theta}] =\delta^t$.
The empirical characteristic function of the model residuals is given by
\begin{align*}
    \varphi_n(t)=\frac{1}{n}\sum_{j=1}^ne^{it\widehat{\theta}_{\varepsilon_j}}.
\end{align*}
A class of test statistics proposed in \cite{rao2019class}, modified to test the hypothesis \eqref{hipoteza}, now has the form
\begin{align}T_n=n\sum_{t=0}^{\infty}|\varphi_n(t)-\varphi(t,\widehat{\delta})|^2\omega(t)=n\sum_{t=0}^{\infty}|\varphi_n(t)-\widehat{\delta}^t|^2\omega(t),\end{align}
where $\omega(\cdot)$ denotes a weight function that satisfies Assumption C from Appendix \ref{asumptionC}. Typically, it is a probability mass function on the non-negative integers with a finite second moment, such as the mass function of the Poisson distribution.

\section{Asymptotic properties}\label{sec: asymptotic}

Here we present the null distribution of the test statistic $T_n$ under the assumption that the errors in the model follow the wrapped Cauchy distribution with an unknown parameter. For this purpose, following \cite{rao2019class}, we introduce $l^2_\omega$ - the separable Hilbert space of all infinite sequences $z = (z_0,z_1,...)$ of complex numbers such that $\sum\limits_{t\ge 0}|z_t|^2 \omega(t)<\infty$, where the inner product is defined as $\langle z,u\rangle_\omega=\sum\limits_{t\geq 0}z_t\overline{u_t} \omega(t)$ with $z,u\in l^2_\omega$ and the norm is denoted by $||\cdot||_{\omega}$. 
%Then statistic $T_n$ can be represented as $T_n=||Z_n||_w$
With this notation, the test statistic may be written as 
$T_n=\|G_n\|_\omega^2$, where 
$G_n(t)=\sqrt{n}(\varphi_n(t)-\varphi(t, \widehat{\delta})).$

Although we assumed in the previous section that all parameters were obtained using the MLE, here we derive the null distribution of $T_n$ under somewhat more general conditions. In particular, we require that Assumptions A, B, and C of the appendix \ref{app_assumptions} are satisfied, while it is clear that they are fulfilled for MLE (see, e.g., \cite{hogg2019introduction}). We have the following result.

\begin{theorem}
    If the null hypothesis is true, and under the assumptions A, B, and C given in the Appendix \ref{app_assumptions}, there exists a zero-mean Gaussian element $\mathcal{G}\in l_\omega^2$ with covariance kernel {$K_{\mathcal{G}}(s, t)$  defined in (\ref{covkernel})  } such that 
    $$T_{n} \stackrel{{\cal L}}{\longrightarrow} \|\mathcal{G}\|_{\omega}^2.$$
\end{theorem}
The proof of the theorem is postponed to the Appendix \ref{proof}.
As a corollary, %it can be shown using standard techniques that 
the limiting null distribution of the test statistic $T_n$ can be represented as an infinite weighted sum of independent chi-squared variables. 
However, these weights are the eigenvalues of the integral operator associated with the kernel $K_{\mathcal{G}}(\cdot,\cdot)$ defined in (\ref{covkernel}), and computing these eigenvalues is a very difficult task. For more details, we refer to \cite{ebner2024eigenvalues}.
%However, it's important to note that these weights, and thus the distribution, depend on the model parameters. 
Therefore, it is recommended to approximate the $p$-values using bootstrap techniques. For Details, see Appendix \ref{app:class-boot}.

\section{Empirical study}\label{sec: powers}
In this section, we focus on the power performance of $T_n$ with a weight function equal to the probability mass function of the Poisson law with mean $\lambda\in\{0.3,0.5,1\}$. The choice of $\lambda$ follows the results of \cite{rao2019class}.
For these choices of $\lambda$, the probability mass is concentrated on small integer values and down-weights the higher-order terms, which are known to be more susceptible to the periodic behavior inherent in the empirical characteristic function.

We consider the model \eqref{model} with $(\beta_0,\beta_1)\in\{(e^{i\frac{\pi}{4}},0.9e^{i\frac{\pi}{6}}),(e^{i\frac{\pi}{4}}, 0.1e^{i\frac{\pi}{6}}), (e^{i\frac{3\pi}{4}},0.1e^{i\frac{\pi}{6}})\}.$  The angles of the covariates $\{\theta_{x_j},\;j=1,...,n\}$  in the study are generated from a uniform $U(0,2\pi)$ distribution.

The idea of applying tests originally for i.i.d. data to model residuals is used here for some other goodness-of-fit tests for circular data. However, we have no results on the validity of these procedures. In particular, we consider
\begin{itemize}

\item  Kuiper's test with the test statistic
$$K_n=\underset{1\leq j \leq n}{\max}
\Big\{U_{(j)}-\frac{j-1}{n}\Big\}+
\underset{1\leq j \leq n}{\max}
\Big\{\frac{j}{n}-U_{(j)}\Big\},$$ 
\item  Watson's test with test statistic
$$W_n=\frac{1}{12n}+\sum\limits_{j=1}^n \Big(\Big(U_{(j)}-\frac{2j-1}{2n}\Big)-\Big(\overline{U}-\frac{1}{2}\Big)\Big)^2,$$
\end{itemize}
where $U_{(1)}\leq \ldots \leq U_{(n)}$ are the order statistics of $U_j=F(\theta_{\varepsilon_j},\widehat{\delta}), j=1,\ldots,n$, $\overline{U}=\frac{1}{n}\sum\limits_{j=1}^nU_j$ and $F(\cdot,\delta)$ denotes the distribution function of $WC(\delta)$.

Since the test statistics are not distribution-free, even asymptotically, we approximate their powers and empirical sizes using a warp-speed bootstrap algorithm given in \cite{giacomini2013warpspeed} with $B=10000$ replicates to reduce computing time. We consider the sample sizes $n=25,50$ and 100, and the following alternative distributions for the model innovations:

\begin{itemize}
    \item wrapped normal (WN$(\rho)$)
    \item von Mises distribution (VM$(\kappa)$)
    \item cardioid distribution (C$(\rho)$)
    \item Cartwright's power-of-cosine distribution (CW($\zeta$))
    \item Jones-Pewsey distribution (JP($\kappa,\psi$))
    \item Batschelet distribution (Ba($\kappa,\nu$)).
\end{itemize}
The detailed specification of the alternatives is given in the Appendix \ref{alternatives}.
For all computations in this and the following section, we used the statistical computing environment \texttt{R} \citep{R2024}, together with the 
%For the generation of data from mentioned alternatives we use 
R packages \texttt{circular} \citep{circularR} and \texttt{cylcop} \citep{cylcopR}.
%\cite{circular_pack})
 
 Tables \ref{tab: size109}-\ref{tab: size301} present the tests' sizes for significance levels $\alpha \in \{0.01, 0.05, 0.1\}$, while Tables \ref{tab: powersall_109}-\ref{tab: powersall_301} provide the empirical powers for $\alpha = 0.05$, both evaluated under various model parameters.
 
 We can see that for larger sample sizes all tests are well-calibrated.
 %, while for smaller $n$, $T_n$ for $\lambda=1$ (denoted by $T_n^{(1)}$) exhibits liberal behavior.
Considering empirical powers, we can conclude that $T_n$ is more powerful than $W_n$ and $K_n$, with a tendency for the difference between them to decrease with increasing sample size. It is also important to note the dominance of $T_n$ over close alternatives. When $W_n$ and $K_n$ are compared, $W_n$ is usually more powerful.

\begin{table}[!htb]
\centering
\setlength{\tabcolsep}{1pt} %reduce column width
\begin{tabular}{ccccccc|ccccc|ccccc}
\noalign{\smallskip}&\multicolumn{6}{c}{$n=25$}
&\multicolumn{5}{c}{$n=50$}&\multicolumn{5}{c}{$n=100$}\\
\noalign{\smallskip}\hline\noalign{\smallskip}
$\alpha$&Alternative&$T_n^{(0.3)}$&$T_n^{(0.5)}$&$T_n^{(1)}$&$K_n$&$W_n$&$T_n^{(0.3)}$&$T_n^{(0.5)}$&$T_n^{(1)}$&$K_n$&$W_n$&$T_n^{(0.3)}$&$T_n^{(0.5)}$&$T_n^{(1)}$&$K_n$&$W_n$ \\
\noalign{\smallskip}\hline\noalign{\smallskip}
&$\mathcal{WC}(0.1)$&1 & 1 & 1 & 1 & 1&1 & 1 & 1 & 1 & 1 &1 & 1 & 1 & 1 & 1\\
0.01&$\mathcal{WC}(0.5)$ & 1 & 1 & 1 & 1 & 1 & 1 & 1 & 1 & 1 & 1 & 1 & 1 & 1 & 1 & 1\\
&$\mathcal{WC}(0.9)$&1 & 1 & 1 & 1 & 1&1 & 1 & 1 & 1 & 1&1 & 1 & 1 & 1 & 1\\
\noalign{\smallskip}\hline\noalign{\smallskip}
& $\mathcal{WC}(0.1)$ & 4 & 5 & 6 & 5 & 5 & 4 & 5 & 6 & 5 & 5 & 5 & 5 & 5 & 5 & 5\\
0.05 & $\mathcal{WC}(0.5)$ &5 & 5 & 5 & 4 & 5 & 4 & 4 & 5 & 5 & 5 & 5 & 5 & 5 & 5 & 5\\
& $\mathcal{WC}(0.9)$ & 6 & 5 & 5 & 5 & 5 & 6 & 6 & 6 & 5 & 5 & 5 & 5 & 5 & 5 & 5\\
\noalign{\smallskip}\hline\noalign{\smallskip}
 & $\mathcal{WC}(0.1)$& 9 & 10 & 12 & 10 & 10 & 9 & 10 & 11 & 11 & 10 & 9 & 10 & 10 & 10 & 10\\
0.1 & $\mathcal{WC}(0.5)$ & 10 & 10 & 10 & 10 & 9 & 9 & 9 & 10 & 9 & 9 & 10 & 11 & 11 & 10 & 10\\
 & $\mathcal{WC}(0.9)$&11 & 11 & 11 & 10 & 9 & 11 & 11 & 11 & 10 & 10 & 10 & 10 & 10 & 10 & 10\\
\hline
\end{tabular}
\captionsetup{justification=centering}
\caption{Empirical sizes for different sample sizes at significance level $\alpha$; $\beta_0=e^{i\frac{\pi}{4}}, \beta_1=0.9e^{i\frac{\pi}{6}}$}
\label{tab: size109}
\end{table}

\begin{table}[!htb]
\centering
\setlength{\tabcolsep}{1pt} %reduce column width
\begin{tabular}{ccccccc|ccccc|ccccc}
\noalign{\smallskip}&\multicolumn{6}{c}{$n=25$}
&\multicolumn{5}{c}{$n=50$}&\multicolumn{5}{c}{$n=100$}\\
\noalign{\smallskip}\hline\noalign{\smallskip}
$\alpha$&Alternative&$T_n^{(0.3)}$&$T_n^{(0.5)}$&$T_n^{(1)}$&$K_n$&$W_n$&$T_n^{(0.3)}$&$T_n^{(0.5)}$&$T_n^{(1)}$&$K_n$&$W_n$&$T_n^{(0.3)}$&$T_n^{(0.5)}$&$T_n^{(1)}$&$K_n$&$W_n$ \\
\noalign{\smallskip}\hline\noalign{\smallskip}
&$\mathcal{WC}(0.1)$&1 & 1 & 1 & 1 & 1&1 & 1 & 1 & 1 & 1 &1 & 1 & 1 & 1 & 1\\
0.01&$\mathcal{WC}(0.5)$ & 1 & 1 & 1 & 1 & 1 & 1 & 1 & 1 & 1 & 1 & 1 & 1 & 1 & 1 & 1\\
&$\mathcal{WC}(0.9)$&1 & 1 & 1 & 1 & 1&1 & 1 & 1 & 1 & 1&1 & 1 & 1 & 1 & 1\\
\noalign{\smallskip}\hline\noalign{\smallskip}
&$\mathcal{WC}(0.1)$&4&5&6&5&5&4&5&5&5&5&4&5&5&5&5\\
0.05&$\mathcal{WC}(0.5)$&4&5&5&5&5&4&5&5&5&5&5&5&5&5&5\\
&$\mathcal{WC}(0.9)$&6&6&6&6&5&6&6&6&6&5&6&5&5&5&5\\
\noalign{\smallskip}\hline\noalign{\smallskip}
 & $\mathcal{WC}(0.1)$& 8 & 10 & 12 & 10 & 10 & 9 & 10 & 12 & 10 & 10 & 8 & 9 & 10 & 10 & 10\\
0.1 & $\mathcal{WC}(0.5)$ & 9 & 9 & 10 & 10 & 9 & 10 & 10 & 10 & 10 & 10 & 10 & 10 & 11 & 11 & 10\\
 & $\mathcal{WC}(0.9)$& 12 & 12 & 11 & 10 & 10 & 11 & 11 & 11 & 10 & 10 & 11 & 11 & 11 & 10 & 10\\
\hline
\end{tabular}
\captionsetup{justification=centering}
\caption{Empirical sizes for different sample sizes at significance level $\alpha$; $\beta_0=e^{i\frac{\pi}{4}}, \beta_1=0.1e^{i\frac{\pi}{6}}$}
\label{tab: size101}
\end{table}

\begin{table}[!htb]
\centering
\setlength{\tabcolsep}{1pt} %reduce column width
\begin{tabular}{ccccccc|ccccc|ccccc}
\noalign{\smallskip}&\multicolumn{6}{c}{$n=25$}
&\multicolumn{5}{c}{$n=50$}&\multicolumn{5}{c}{$n=100$}\\
\noalign{\smallskip}\hline\noalign{\smallskip}
$\alpha$&Alternative&$T_n^{(0.3)}$&$T_n^{(0.5)}$&$T_n^{(1)}$&$K_n$&$W_n$&$T_n^{(0.3)}$&$T_n^{(0.5)}$&$T_n^{(1)}$&$K_n$&$W_n$&$T_n^{(0.3)}$&$T_n^{(0.5)}$&$T_n^{(1)}$&$K_n$&$W_n$ \\
\noalign{\smallskip}\hline\noalign{\smallskip}
&$\mathcal{WC}(0.1)$&1 & 1 & 1 & 1 & 1&1 & 1 & 1 & 1 & 1 &1 & 1 & 1 & 1 & 1\\
0.01&$\mathcal{WC}(0.5)$ & 1 & 1 & 1 & 1 & 1 & 1 & 1 & 1 & 1 & 1 & 1 & 1 & 1 & 1 & 1\\
&$\mathcal{WC}(0.9)$&1 & 1 & 1 & 1 & 1&1 & 1 & 1 & 1 & 1&1 & 1 & 1 & 1 & 1\\
\noalign{\smallskip}\hline\noalign{\smallskip}
&$\mathcal{WC}(0.1)$&4&5&6&5&4&4&5&6&5&5&4&4&5&5&4\\
0.05&$\mathcal{WC}(0.5)$&4&4&5&5&5&5&5&5&5&5&5&5&5&5&5\\
&$\mathcal{WC}(0.9)$&5&5&6&5&5&6&6&6&5&5&5&5&5&5&5\\
\noalign{\smallskip}\hline\noalign{\smallskip}
 & $\mathcal{WC}(0.1)$& 8 & 9 & 12 & 10 & 9 &  8 & 10 & 11 & 9 & 10 & 8 & 9 & 10 & 10 & 9\\
0.1 & $\mathcal{WC}(0.5)$ & 9 & 9 & 10 & 10 & 9 & 10 & 10 & 10 & 10 & 10 & 10 & 10 & 11 & 10 & 10\\
 & $\mathcal{WC}(0.9)$& 11 & 11 & 10 & 10 & 10 & 11 & 11 & 11 & 10 & 11 & 11 & 11 & 11 & 10 & 10\\
\hline
\end{tabular}
\captionsetup{justification=centering}
\caption{Empirical sizes for different sample sizes at significance level $\alpha$; $\beta_0=e^{i\frac{3\pi}{4}}, \beta_1=0.1e^{i\frac{\pi}{6}}$}
\label{tab: size301}
\end{table}

\begin{table}[!htb]
\centering
\setlength{\tabcolsep}{1pt} %reduce column width
\begin{tabular}{cccccc|ccccc|ccccc}
\noalign{\smallskip}&\multicolumn{5}{c}{$n=25$}
&\multicolumn{5}{c}{$n=50$}&\multicolumn{5}{c}{$n=100$}\\
\noalign{\smallskip}\hline\noalign{\smallskip}
Alternative&$T_n^{(0.3)}$&$T_n^{(0.5)}$&$T_n^{(1)}$&$K_n$&$W_n$&$T_n^{(0.3)}$&$T_n^{(0.5)}$&$T_n^{(1)}$&$K_n$&$W_n$&$T_n^{(0.3)}$&$T_n^{(0.5)}$&$T_n^{(1)}$&$K_n$&$W_n$ \\
\noalign{\smallskip}\hline\noalign{\smallskip}
$\mathcal{WN}(0.5)$ & 4 & 5 & \textbf{6} & 5 & 5 & 14 & 16 & \textbf{18} & 11 & 10 & 45 & \textbf{47} & 46 & 30 & 31\\
$\mathcal{WN}(0.7)$&7&7&\textbf{8}&7&6&45&\textbf{46}&45&29&26&92&\textbf{93}&91&72&68\\
$\mathcal{WN}(0.9)$ & 8& 8& 7&\textbf{12}& 10&\textbf{77}&75&70&56&48&\textbf{97}&\textbf{97}&95&87&84\\
$\mathcal{VM}$(0.9) &4&\textbf{5}&\textbf{5}&\textbf{5}&4&4&\textbf{5}&\textbf{5}&\textbf{5}&\textbf{5}&8&\textbf{9}&\textbf{9}&7&7\\
$\mathcal{VM}$(2) & 5&5&5&\textbf{6}&5&\textbf{25}&\textbf{25}&\textbf{25}&18&17&\textbf{68}&\textbf{68}&65&45&45\\
$\mathcal{VM}$(5) & 7&7&6&\textbf{12}&9&\textbf{71}&68&61&48&43&\textbf{96}&\textbf{96}&94&84&80\\
$\mathcal{VM}$(7) &5&6&6&\textbf{13}&11&\textbf{73}&72&69&55&49&\textbf{96}&\textbf{96}&94&87&83\\
Ca(0.3)&4 &5 &\textbf{7}& 5& 5&5& 6& \textbf{7}& 5& 6&8 &9 &\textbf{10}&  8&  7\\
Ca(0.5)&6 &8 &\textbf{9} &6 &6&26& 29 &\textbf{33}& 18& 17&72& \textbf{75}& \textbf{75}& 51& 48\\
CW(0.5)& 8 &8 &\textbf{9}& 8& 7&54 &\textbf{55 }&\textbf{55}& 33& 29&\textbf{94} &\textbf{94}& \textbf{94} &78 &75\\
CW(1) &7 &8& \textbf{9}& 6& 6& 25 &29& \textbf{33} &18 &18&72& \textbf{75} &\textbf{75 }&51& 48\\
JP(2,0)&5 &5& 4& \textbf{6}& 5&\textbf{27} &26 &25 &18 &17&\textbf{67}& \textbf{67}& 64 &45& 45\\
JP(2,1)&6 &7 &\textbf{9} &6& 6&21 &25& \textbf{28}& 16& 16&63 &\textbf{66} &\textbf{66} &43& 41\\
JP(2,1.5)&4& 5& \textbf{7}& 6 &5&13 &16& \textbf{19}& 11 &10&43& 47 &\textbf{49} &30& 29\\
Ba(3,0.5)&4 &4 &4 &\textbf{8}& 7&\textbf{45}& \textbf{45}& 41& 33& 30&\textbf{83} &82& 78& 65 &62\\
Ba(3,1)&2 & 3 & 3& \textbf{10} & 8&37 &\textbf{39} &\textbf{39}& 36 &33&77& \textbf{78} &\textbf{78} &68 &65\\
%Ba(3,2)&1& 1& 1& \textbf{9}& 7& 9& 13 &22& \textbf{31}& 29&20& 28& 48& \textbf{66} &63\\
\end{tabular}
\captionsetup{justification=centering}
\caption{Empirical {size (first 3 lines) and} power (expressed as a percentage) for different sample sizes at significance level 0.05; $\beta_0=e^{i\frac{\pi}{4}}, \beta_1=0.9e^{i\frac{\pi}{6}}$ }
\label{tab: powersall_109}
\end{table}

\begin{table}[!htb]
\centering
\setlength{\tabcolsep}{1pt} %reduce column width
\begin{tabular}{cccccc|ccccc|ccccc}
\noalign{\smallskip}&\multicolumn{5}{c}{$n=25$}
&\multicolumn{5}{c}{$n=50$}&\multicolumn{5}{c}{$n=100$}\\
\noalign{\smallskip}\hline\noalign{\smallskip}
Alternative&$T_n^{(0.3)}$&$T_n^{(0.5)}$&$T_n^{(1)}$&$K_n$&$W_n$&$T_n^{(0.3)}$&$T_n^{(0.5)}$&$T_n^{(1)}$&$K_n$&$W_n$&$T_n^{(0.3)}$&$T_n^{(0.5)}$&$T_n^{(1)}$&$K_n$&$W_n$ \\
\noalign{\smallskip}\hline\noalign{\smallskip}
$\mathcal{WN}(0.5)$ &5 &6 & \textbf{8}& 7& 6& 20& 22& \textbf{23}& 15 & 15 & 54 & \textbf{56} & 55 & 36 & 38\\
$\mathcal{WN}(0.7)$& 12 & 12 & \textbf{13}& 12 & 11 &  \textbf{54} & \textbf{54}& 52& 37& 34 & \textbf{97} & \textbf{97}& \textbf{97}& 81 &77\\
$\mathcal{WN}(0.9)$ & 14& 14& 12& \textbf{17}& 15 & \textbf{82}& 80& 76& 60& 52 &  \textbf{100} & \textbf{100} & \textbf{100} & 99 & 95\\
$\mathcal{VM}$(0.9) &3 &4 &\textbf{5}& 4& \textbf{5}&5 &\textbf{6}& \textbf{6}& \textbf{6} &\textbf{6}&11& \textbf{12}& \textbf{12}&  9& 10\\
$\mathcal{VM}$(2) &7 &7& 7& \textbf{8}& \textbf{8}&\textbf{30} &29& 28& 21& 21&\textbf{78}& 77& 75& 56& 56\\
$\mathcal{VM}$(5) & 12& 11 &10& \textbf{16}& 14& \textbf{76} &74& 68& 52& 47&\textbf{100} &\textbf{100} &\textbf{100}  &96 & 92\\
$\mathcal{VM}$(7) &9 &10 &10& \textbf{16} &14&\textbf{77}& \textbf{77} &73 &56 &49&  \textbf{100}& \textbf{100} &\textbf{100}&  98  &94\\
Ca(0.3)& 4& 5& \textbf{6}&5 &5&6 &7& \textbf{8} &6 &7&11 &11& \textbf{12}&  9& 10\\
Ca(0.5)& 8 & 9 &\textbf{12}&  8 & 8&34 &37 &\textbf{38}& 24& 24&81& \textbf{83}& 82& 60& 59\\
CW(0.5)&  13& 14& \textbf{16}& 12& 11&\textbf{62} &\textbf{62} &60& 40& 36&98& \textbf{99}& 98& 88& 83\\
CW(1) &  8 & 9 &\textbf{12} & 8&  8&34 &37& \textbf{38} &24& 24& 81 &\textbf{83} &82 &60& 59\\
JP(2,0)&6& 6& 6 &\textbf{8}& \textbf{8}& \textbf{30}& \textbf{30}& 29& 21& 20& \textbf{77} &\textbf{77}& 75& 55 &55\\
JP(2,1)& 7&  8 &\textbf{11}&  8  &8&  27& 30& \textbf{31} &20 &20& 72& \textbf{75}& \textbf{75} &53& 53\\
JP(2,1.5)&5 &7 &\textbf{9}& 7 &7&  17 &20& \textbf{21} &14& 14 & 53& \textbf{56} &\textbf{56}& 39& 40\\
%Ba(2,1)&\\
Ba(3,0.5)&  7 & 7 & 7 &\textbf{12}& 11 &\textbf{50} &49& 45& 35 &31 &\textbf{95}& \textbf{95}& 94 &81 &79\\
Ba(3,1)&  4 & 4 & 5 &\textbf{12} &11& \textbf{42}& \textbf{42} &\textbf{42}& 37& 32& 93& \textbf{94} &\textbf{94} &83 &79\\
%Ba(3,2)&  1 &1& 3& \textbf{9}& 8& 10& 14 &22& \textbf{29}& 27& 50 &60 &\textbf{78} &76& 72\\
\end{tabular}
\captionsetup{justification=centering}
\caption{Empirical powers (expressed as a percentage) for different sample sizes at significance level 0.05; $\beta_0=e^{i\frac{\pi}{4}}, \beta_1=0.1e^{i\frac{\pi}{6}}$}
\label{tab: powersall_101}
\end{table}

\begin{table}[!htb]
\centering
\setlength{\tabcolsep}{1pt} %reduce column width
\begin{tabular}{cccccc|ccccc|ccccc}
\noalign{\smallskip}&\multicolumn{5}{c}{$n=25$}
&\multicolumn{5}{c}{$n=50$}&\multicolumn{5}{c}{$n=100$}\\
\noalign{\smallskip}\hline\noalign{\smallskip}
Alternative&$T_n^{(0.3)}$&$T_n^{(0.5)}$&$T_n^{(1)}$&$K_n$&$W_n$&$T_n^{(0.3)}$&$T_n^{(0.5)}$&$T_n^{(1)}$&$K_n$&$W_n$&$T_n^{(0.3)}$&$T_n^{(0.5)}$&$T_n^{(1)}$&$K_n$&$W_n$ \\
\noalign{\smallskip}\hline\noalign{\smallskip}
$\mathcal{WN}(0.5)$ &5& 7& \textbf{9}& 7& 7& 20& 22 &\textbf{23} &15 &15 &54 &\textbf{56} &55 &37 &39\\
$\mathcal{WN}(0.7)$&12 &12 &\textbf{13} &\textbf{13} &11 &\textbf{54} &53 &52 &37 &34  &\textbf{97} &\textbf{97} &\textbf{97} &81 &78\\
$\mathcal{WN}(0.9)$ &15 &15 &12 &\textbf{18} &16 & \textbf{82} &81 &77& 60 &53 &\textbf{100}& \textbf{100} &\textbf{100} & 98 & 95
\\
$\mathcal{VM}$(0.9) &4& 4 &\textbf{6}& 5 &5 &5& 6 &\textbf{7}& 6 &6&  11 &\textbf{12} &\textbf{12}& 10& 10 \\
$\mathcal{VM}$(2) &7& 7& 8& \textbf{9}& \textbf{9}& \textbf{31} &30 &29 &22& 22 & \textbf{78} &\textbf{78} &75 &56 &57 \\
$\mathcal{VM}$(5) &  13& 13 &11 &\textbf{17} &15 &\textbf{76}& 74 &68& 51 &47& \textbf{100} &\textbf{100} &\textbf{100} & 96 & 92\\
$\mathcal{VM}$(7)  &11 &12& 12 &\textbf{17}& 15 &\textbf{77}& 76 &73& 56 &49 &  \textbf{100} &\textbf{100} &\textbf{100}&  98 & 94\\
Ca(0.3)&  4 &4& \textbf{6}& \textbf{6}& 5&  7 &8& \textbf{9}& 7& 7& 11& 12& \textbf{13}&  9& 10\\
Ca(0.5)&  9& 11 &\textbf{14}&  9 & 9& 34 &37& \textbf{38} &25& 24 &80 &\textbf{82}& \textbf{82} &60& 59\\
CW(0.5)& 14& 15 &\textbf{17}& 13 &11&  \textbf{62} &\textbf{62} &61& 39 &36& 98& \textbf{99} &98& 87& 83\\
CW(1) & 9 &11 &\textbf{14}&  9 & 9& 34 &37& \textbf{38} &25& 24 &80& \textbf{82}& \textbf{82}& 60& 59\\
JP(2,0)&  7 &6 &7 &\textbf{9} &\textbf{9} &\textbf{30} &29 &29 &21 &21 &\textbf{77} &\textbf{77} &75 &55 &55\\
JP(2,1)& 8 & 9 &\textbf{12} & 9 & 9& 28 &31& \textbf{32} &20& 20& 72 &\textbf{75} &\textbf{75}& 53& 54\\
JP(2,1.5)&  6&  8 &\textbf{11} & 8 & 8 &19 &\textbf{21} &\textbf{21}& 14 &15&  53& \textbf{56} &\textbf{56} &38 &39\\
%Ba(2,1)&\\
Ba(3,0.5)&8 & 7 & 7 &\textbf{13}& 12 &\textbf{50} &49 &46 &35 &32 & \textbf{95}& \textbf{95} &94& 82 &79\\
Ba(3,1)&5 & 5 & 6 &\textbf{12} &11& 41 &\textbf{43}& \textbf{43} &36 &32 & 93 &\textbf{94} &\textbf{94} &83& 78\\
%Ba(3,2)& 1  &2&  3 &\textbf{10}&  9 & 10 &14& 22 &\textbf{29}& 27 & 50& 60& \textbf{78} &76& 71\\
\end{tabular}
\captionsetup{justification=centering}
\caption{Empirical powers (expressed as a percentage) for different sample sizes at significance level 0.05; $\beta_0=e^{i\frac{3\pi}{4}}, \beta_1=0.1e^{i\frac{\pi}{6}}$}
\label{tab: powersall_301}
\end{table}

\section{Real data examples}\label{sec: realdata} 

In this section, we fit the circular-circular regression model in (\ref{model}) to three small datasets from the literature and a larger dataset of wind data from the Black Forest in southwestern Germany and test whether the assumed model is a reasonable choice, i.e. we test the hypothesis in (\ref{hipoteza}). All $p$-values are obtained using $B=10000$ bootstrap replicates.

\textbf{Example 1:} The wind direction at 6 a.m. and 12 p.m. was measured every day for 21 consecutive days at a weather station in Milwaukee \cite{johnson1977measures}. The dataset is shown in Table \ref{tab: datasetWind}. We use model (\ref{model}) to regress the wind direction at noon to the wind direction at 6 am. 
%The maximum likelihood estimates of the parameters and further discussion can be found in \cite{kato2008circular}.
The maximum likelihood estimates are 
$\hat{\theta}_0=1.27, \hat{r}=0.53, \hat{\theta_1}=2.59$, and $\hat{\delta}=0.55$. 
The results of the different goodness-of-fit tests shown in the first row of Table \ref{tab: pvaluesRealData} indicate that the data are consistent with the model (\ref{model}).

\begin{table}[!htb]
\centering
\begin{tabular}{lccccccccccc}
\hline\hline\noalign{\smallskip}
\textbf{6 a.m.} & 356 & 97.2 & 211 & 232 & 343 & 292 & 157 & 302 & 335 & 302 &324\\&84.6&324& 340& 157& 238& 254& 146& 232& 122& 329\\
\noalign{\smallskip}\hline\noalign{\smallskip}
\textbf{12 a.m.} &119& 162& 221& 259& 270& 28.8& 97.2& 292& 39.6& 313&94.2\\
&45& 47& 108& 221& 270& 119& 248& 270& 45& 23.4\\
\hline \noalign{\smallskip}
\end{tabular}
 \caption{Dataset 1 - Wind direction (in degrees)}\label{tab: datasetWind}
\end{table}

\textbf{Example 2:} In a medical experiment, several measurements were taken on 10 medical students several times a day over several weeks \cite{downs1974rotational}. The estimated peak times (converted into angles $\theta$ and $\phi$) for two consecutive measurements of diastolic blood pressure are given in Table \ref{tab: datasetBloodPreasure}.
The estimated parameters are 
$\hat{\theta}_0=0.01, \hat{r}=0.06, \hat{\theta_1}=5.22$, and $\hat{\delta}=0.97$. 
The value of $\hat{\delta}$ is close to 1, indicating that covariates and responses are correlated almost without error. A look at the fitted values confirms this, see the third row in Table \ref{tab: datasetBloodPreasure}. As expected, the validity of model (\ref{model}) is supported by the results of the goodness-of-fit tests given in the second row of Table \ref{tab: pvaluesRealData}.

\begin{table}[!htb]
\centering
\begin{tabular}{lcccccccccccc}
\hline\hline\noalign{\smallskip}
$\boldsymbol{\theta}$ & 30& 15& 11& 4& 348& 347& 341& 333& 332& 285\\
\noalign{\smallskip}\hline\noalign{\smallskip}
$\boldsymbol{\phi}$ &25& 5& 349& 358& 340& 347& 345& 331& 329& 287\\
\noalign{\smallskip}\hline\noalign{\smallskip}
$\boldsymbol{\widehat{\phi}}$ &24& 9&   5& 359& 344& 343& 337& 330& 329& 287 \\
\hline \noalign{\smallskip}
\end{tabular}
 \caption{Dataset 2 - Blood pressure}\label{tab: datasetBloodPreasure}
\end{table}

\textbf{Example 3:} This data set contains 38 phase or peak expression times of
synchronized circadian-related genes common to heart and liver tissue in vivo \cite{liu2006phase}. The phase angles (in radians) of the circadian-related transcripts in the heart and liver are given in Table \ref{tab: datasetPeakTimes}.
The maximum likelihood estimates are 
$\hat{\theta}_0=0.11, \hat{r}=0.27, \hat{\theta_1}=2.36$, and $\hat{\delta}=0.61$. 
Again, the $p$-values shown in the third row of Table \ref{tab: pvaluesRealData} support the validity of the regression model in (\ref{model}).

\begin{table}[!htb]
\centering
\begin{tabular}{lcccccccccccc}
\hline\hline\noalign{\smallskip}
$\boldsymbol{\theta}$  & 0.12& 0.27&0.29& 0.3& 0.31& 0.34& 0.35&0.58& 0.62& 1.6\\&2.35&2.62&2.83
                        & -3.06& -2.86&-2.77& -2.69& -2.57& -2.56& -2.45\\& -2.43 &-2.37&-2.18& -2.16&-2.04&-1.61
                        & -1.32& -1.22& -0.84& -0.77\\& -0.38& -0.36&-0.26& -0.19& -0.18& -0.13& -0.12& -0.02\\
\noalign{\smallskip}\hline\noalign{\smallskip}
$\boldsymbol{\phi}$  &0.61& 0.95& -2.85& 0.67& -0.13& 0.08& 2.67& 1.72& 1.45& 1.59\\& -2.51&-2.92& 1.42&
                    2.74&2.88& -3.01& -2.69& 3.05&-2.35& 2.68\\& -2.86&-2.51&2.69& -2.11& -1.48& -2.06&-2.63&-1.49& -0.83& 0.86\\& 0.26& 1.5& 1.03& 0.33&  -1.15& -0.21& -0.55&0.91\\
\hline \noalign{\smallskip}
\end{tabular}
 \caption{Dataset 3 - Peak expression times}\label{tab: datasetPeakTimes}
\end{table}

\begin{table}[!htb]
\centering
\begin{tabular}{lccccc}
\hline\noalign{\smallskip}
Dataset&$T_n^{(0.3)}$&$T_n^{(0.5)}$&$T_n^{(1)}$&$K_n$&$W_n$\\
\noalign{\smallskip}\hline\noalign{\smallskip}
wind direction&0.18 &0.12& 0.06& 0.39& 0.56\\
blood preasure&0.58 &0.59& 0.62& 0.73& 0.70\\
peak times    &0.36 &0.33& 0.30& 0.77& 0.77\\
\hline \noalign{\smallskip}
\end{tabular}
\captionsetup{justification=centering}
 \caption{Empirical $p$-values for data examples in Section \ref{sec: realdata}}\label{tab: pvaluesRealData}
\end{table}

\smallskip
%\section{Application to Black Forest wind data} \label{sec:blackforest}
\textbf{Example 4:} \label{sec:blackforest}
The wind conditions at a potential site are critical to the design of wind turbines. When considering a site, the local wind conditions are analyzed using calculations based on reference values or on-site measurements. The second method is expensive and only available for relatively short periods. Reference values are usually available free of charge for long periods, but the reference sites are more or less far away and the transferability must be quantified.

\begin{figure}
\centering
\includegraphics[width=0.9\textwidth]{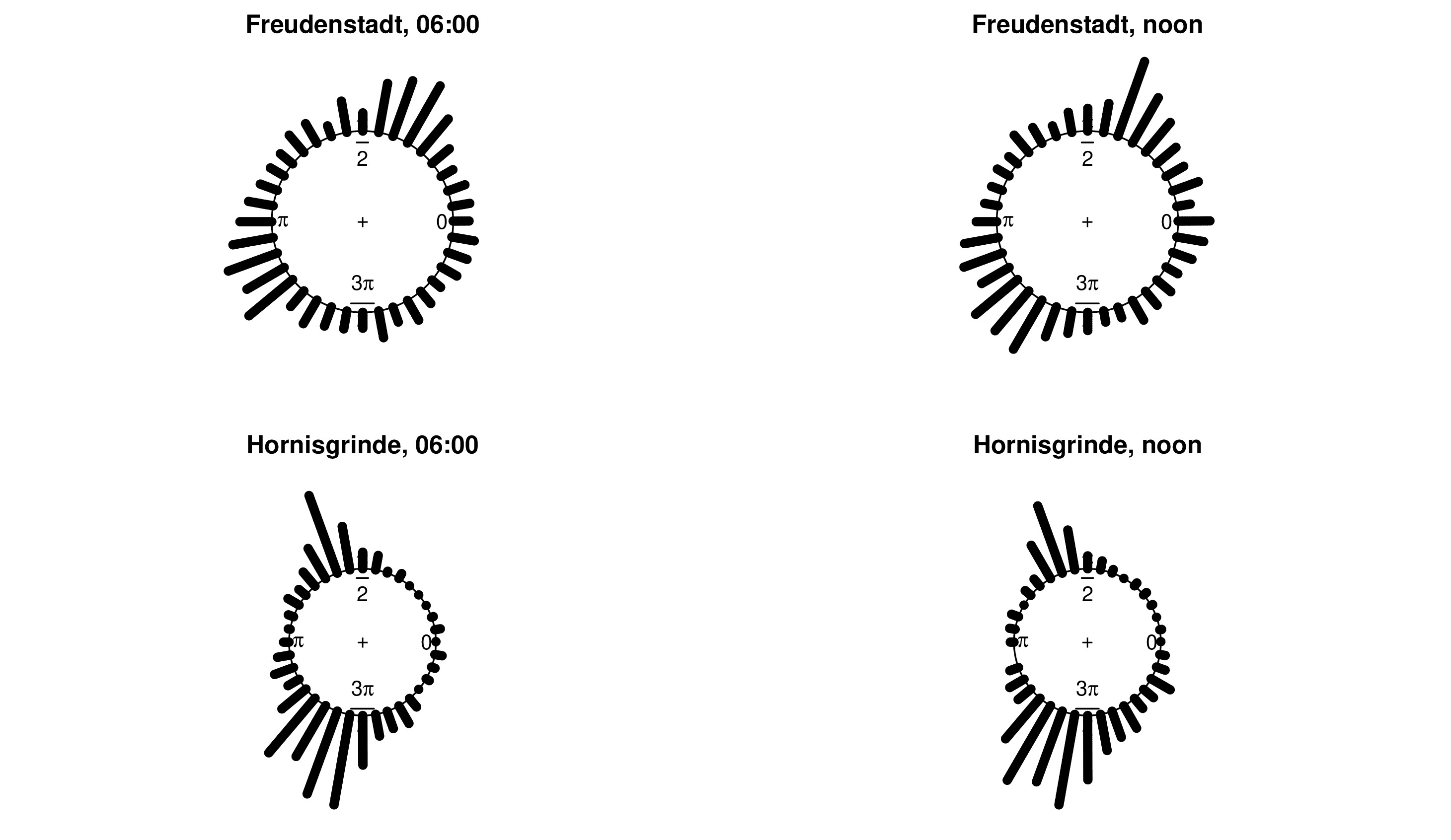}
\captionsetup{justification=centering}
\caption{Stack plot of the wind directions at the two sites, measured at 06:00 and noon over 9 years from 2015 to 2023.} \label{fig-bf-wind-data}
\end{figure}

We consider datasets from the German Weather Service, available at \url{https://opendata.dwd.de/climate_environment/CDC/observations_germany/climate/hourly/wind/}.
The accuracy of the wind direction measurement is 10 degrees, so all values are multiples of 10. We use data from two nearby stations in the Black Forest in southwestern Germany, Freudenstadt (fr) and  Hornisgrinde (ho), at a distance of about 22 km. We regress the values at a station as the potential site on the nearby "reference station" as a proxy for what is done in real projects.
The datasets contain wind directions measured at 6:00 am and noon on Wednesdays over 9 years, from 2015/01/07 to 2023/12/27. After removing a small number of missing values, the sample size of each measured variable is 463. The four resulting data sets are fr06, fr12, ho06 and ho12.
Although the lag between consecutive observations is one week, the data are time series. Therefore, we computed autocorrelations of the four data sets, using a correlation coefficient for angular variables proposed in \cite{JaSa1988}. The lag-one autocorrelations for the four datasets are $0.093, 0.019, -0.003$ and $0.113$, and the $p$-values for corresponding tests of the hypothesis that the autocorrelation is zero are $0.043, 0.686, 0.946$ and $0.017$. We also computed autocorrelations for lags 2 through 5 (corresponding to 2-5 weeks), which yielded two values between 0.1 and 0.2; the remaining values are below 0.1. Out of a total of 20 tests, only one is significant at the 0.01 level. In summary, the serial correlation within the time series is low and does not invalidate the subsequent analysis.
Figure \ref{fig-bf-wind-data} shows stack plots of the wind directions at the two sites.
%Stuttgart-Echterdingen (st1) / Stuttgart-Schnarrenberg (st2): distance about 15 km, dataset data.st: n=468 

We now regress the values at the Hornisgrinde station at noon (ho12) on the corresponding values at the Freudenstadt station (fr12), and repeat the same with the data at 06:00.

As further examples, we regress the wind direction at noon on the wind direction at 6 am as in Example 1 in Section \ref{sec: realdata} to mimic a forecast scenario. The estimated parameters of the 4 regression models are given in Table \ref{tab:parametersBlackForest}.

% latex table generated in R 4.3.2 by xtable 1.8-4 package
% Fri Jun  7 15:52:22 2024
\begin{table}[ht]
\centering
\begin{tabular}{rr|rrrrrrrr}
  \hline
$x$ & $y$ & $n$ & $\hat{\theta}_0$ & $\hat{\theta}_1$ & $r$ & $\delta$ & $\hat{\mu}_{y.\pi/4}$ & $\hat{\mu}_{y.3\pi/4}$ \\   \hline
fr12 & ho12 & 463 & 0.59 & 3.70 & 0.36 & 0.65 & 1.62 & 3.57 \\ 
fr06 & ho06 & 463 & 1.37 & 2.65 & 0.51 & 0.54 & 3.20 & 3.93 \\ 
ho06 & ho12 & 463 & 6.18 & 4.79 & 0.24 & 0.75 & 0.25 & 2.64 \\ 
fr06 & fr12 & 463 & 0.28 & 0.36 & 0.07 & 0.45 & 1.02 & 2.51 \\ 
\hline \noalign{\smallskip}
\end{tabular}
 \caption{Estimated parameters of the different regression models using the Black Forest wind data.}
 \label{tab:parametersBlackForest}
\end{table}

\begin{figure}
\centering
\includegraphics[width=0.7\textwidth]{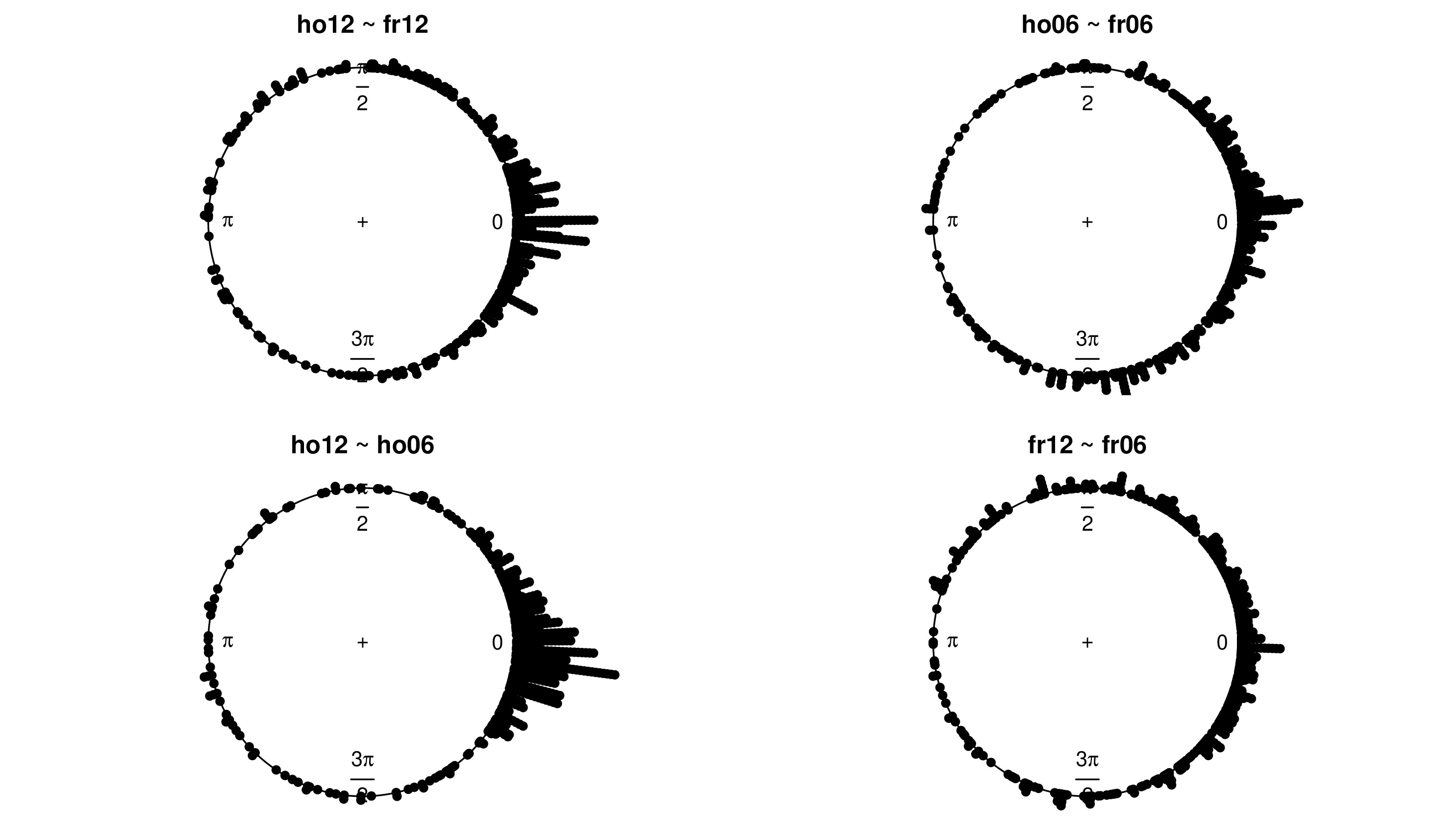}
\captionsetup{justification=centering}
\caption{Stack plot of the residuals $\widehat{\theta}_{\varepsilon}$ of the different regression models.} \label{fig-bf-wind-residuals}
\end{figure}

The conditional distribution of $Y|x$ in model (\ref{model}) is a wrapped Cauchy distribution $WC(\mu_{y.x},\delta_{Y|x})$ with  
\begin{align*}
    \mu_{y.x} = \arg\left( \beta_0 \frac{x+\beta_1}{1+\overline{\beta}_1x} \right), \quad
    \delta_{Y|x} = \exp(i\mu_{y.x}) \delta
\end{align*}
(see \citet[p. 638]{kato2008circular}). Thus, the circular variance takes the constant value $1-\delta$, while the circular mean is $\mu_{y.x}$. The last two columns of Table \ref{tab:parametersBlackForest} show as an example the fitted $\widehat{\mu}_{y.x}$ for $x=\pi/4$ and $x=3\pi/4$.
Figure \ref{fig-bf-wind-residuals} shows stack plots of the residuals from the 4 regression models.

The results of the goodness-of-fit tests for the full data sets, using $10000$ replications, are shown in the upper panel of Table \ref{tab:pvaluesBlackForest}. Due to the length of the dataset, it is not surprising that the first three models have $p$-values close to zero. However, for the last regression model, the $p$-values are all above 0.05.
In the middle and bottom panel, we have reduced the sample to 200 and 100 observations, roughly corresponding to 2020-23 and 2022-23, respectively. For $n=200$, the second model is still clearly rejected by all tests,
and the $p$-values for the first model are below 0.05 for all tests except $K_n$,
while the $p$-values for model 3 are below 0.05 only for the tests based on $K_n$. 
For $n=100$, the $p$-values are generally higher, and never below 0.03. Again, the $p$-values for model 4, which regresses fr12 on fr06, are higher than for the other models. All in all, the proposed model seems to be a reasonable model for the application under consideration.

\begin{table}[httb]
\centering
\begin{tabular}{rrr|rrrrr}
  \hline
$n$ & $x$ & $y$ &$T_n^{(0.3)}$&$T_n^{(0.5)}$&$T_n^{(1)}$&$K_n$&$W_n$ \\ \hline
    & fr12 & ho12 & 0.00 & 0.00 & 0.00 & 0.00 & 0.00 \\ 
463 & fr06 & ho06 & 0.00 & 0.00 & 0.00 & 0.00 & 0.00 \\ 
    & ho06 & ho12 & 0.01 & 0.01 & 0.01 & 0.01 & 0.04 \\ 
    & fr06 & fr12 & 0.09 & 0.10 & 0.11 & 0.07 & 0.07 \\ 
  \hline
    & fr12 & ho12 & 0.02 & 0.02 & 0.04 & 0.11 & 0.02 \\ 
200 & fr06 & ho06 & 0.00 & 0.00 & 0.00 & 0.00 & 0.00 \\ 
    & ho06 & ho12 & 0.26 & 0.23 & 0.18 & 0.01 & 0.08 \\ 
    & fr06 & fr12 & 0.11 & 0.12 & 0.16 & 0.31 & 0.23 \\ 
  \hline
    & fr12 & ho12 & 0.04 & 0.04 & 0.04 & 0.14 & 0.22 \\ 
100 & fr06 & ho06 & 0.03 & 0.03 & 0.03 & 0.10 & 0.27 \\ 
    & ho06 & ho12 & 0.49 & 0.37 & 0.23 & 0.16 & 0.29 \\ 
    & fr06 & fr12 & 0.98 & 0.98 & 0.97 & 0.93 & 0.91 \\
\hline \noalign{\smallskip}
\end{tabular}
 \caption{Empirical $p$-values for the different regression models using the Black Forest wind data.}
 \label{tab:pvaluesBlackForest}
\end{table}

\section*{Acknowledgements}

The research was done as a part of the bilateral cooperation project Modeling complex data - Selection and Specification between the Federal Republic of Germany and the Republic of Serbia funded by the Federal Ministry of Education and Research of Germany and the Ministry of Science, Technological Development and Innovations of the Republic of Serbia (337-00-19/2023-01/6). The work of K.Halaj is supported by the Ministry of Science, Technological Development and Innovations of the Republic of Serbia (the contract 451-03-65/2024-03/200128), while the work of B. Milo\v sevi\' c and M. Veljovi\' c is supported by the Ministry of Science, Technological Development and Innovations of the Republic of Serbia (the
contract the contract 451-03-66/2024-03/200104). The work of B. Milo\v sevi\' c is also supported by the COST action CA21163 - Text, functional and other
high-dimensional data in econometrics: New models, methods, applications (HiTEc).

\section*{Declarations}

\begin{appendix}

\section{Assumptions}\label{app_assumptions}
\textbf{Assumption A.} \label{asumptionA} Under $H_0$, if $\delta$ denotes the true parameter value, then
$$\sqrt{n}(\widehat{\delta}-\delta)=\frac{1}{\sqrt{n}} \sum_{i=1}^n L_1\left(\theta_{\varepsilon_j} ; \delta\right)+r_{\delta},$$
with {$r_{\delta}=o_p(1)$,}  $\mathbb{E}\{L_1(\theta ; \delta)\}=0$ and $\mathbb{E}\left\{L_1(\theta ; \delta)^2 \right\}<\infty$.\\\\
\textbf{Assumption B.} \label{asumptionB} Under $H_0$, if $u=(\theta_0, \theta_1,r)^T$ denotes the true parameter value, then
$$\sqrt{n}(\widehat{u}-u)=\frac{1}{\sqrt{n}} \sum_{i=1}^n L_2\left(\theta_{\varepsilon_j} ; u\right)+r_u,$$
with {$r_{u}=o{{_p}}(1)$,} $\mathbb{E}\{L_2(\theta ; u)\}=\mathbf{0}$, $\mathbb{E}\left\{L_2(\theta ; u) L_2(\theta ; u)^{\top}\right\}<\infty$, {and $L_2(\cdot;)$ is not identically equal to $\mathbf{0}$.} \\\\
\textbf{Assumption C.} \label{asumptionC} 
 The weight function $\omega$ is nonnegative and $$\sum\limits_{t=0}^{\infty} t^2 \omega(t) < \infty.$$
\allowdisplaybreaks
\section{Proof of Theorem 1}\label{proof}
\begin{proof}
    
%\subsubsection*{Proof of Theorem 1}
We can represent the test statistics as 
\begin{align*}
 T_n & =n \cdot \sum_{t=0}^{\infty}\left|\widehat{\varphi}_n(t)-\varphi(t, \widehat{\delta})\right|^2 \omega(t)=\sum_{t=0}^{\infty}\left|I_n^{(1)}(t)+I_n^{(2)}(t)\right|^2 \omega(t),\end{align*}
where
\begin{align*}
I_n^{(1)}(t) &=\frac{1}{\sqrt{n}} \sum_{j=1}^n\left(\cos \left(t \theta_{\widehat{\varepsilon_j}}\right)+i\sin \left(t \theta_{\widehat{\varepsilon_j}}\right)-\delta^t\right), \qquad
I_n^{(2)}(t) =\sqrt{n}\left(\delta^t-\widehat{\delta}^t\right).
\end{align*}
The main idea is to decompose $I_n^{(1)}(t)$ and $I_n^{(2)}(t)$ into sums of independent and identically distributed (i.i.d.) random variables and negligible remainders. This approach allows us to apply the central limit theorem (CLT) in Hilbert space.
We start with the expansion of $I_n^{(1)}(t)$.

We have
\begin{align*}
I_{n}^{(1)}(t)
&= \frac{1}{\sqrt{n}} \sum_{j=1}^n(\cos (t \theta_{\varepsilon_j})+i\sin (t \theta_{\varepsilon_j})-\delta^t)
\\&+\frac{t}{\sqrt{n}}\sum_{j=1}^n (\theta_{\widehat{\varepsilon_j}}-\theta_{\varepsilon_j})\left(i\cos(t\theta_{\varepsilon_j})-\sin(t\theta_{\varepsilon_j})\right) +\frac{1}{\sqrt{n}}\sum_{j=1}^n \left(R_{1j}(t)+iR_{2j}(t)\right),
\end{align*}
where, for some $\gamma \in [0,1]$, 
\begin{align*}
    R_{1j}(t) = -\frac{1}{2}\cos(\gamma t \theta_{\widehat{\varepsilon_j}}+(1-\gamma)t\theta_{\varepsilon_j})t^2(\theta_{\widehat{\varepsilon_j}}-\theta_{\varepsilon_j})^2,
    \\
     R_{2j}(t) = -\frac{1}{2}\sin(\gamma t \theta_{\widehat{\varepsilon_j}}+(1-\gamma)t\theta_{\varepsilon_j})t^2(\theta_{\widehat{\varepsilon_j}}-\theta_{\varepsilon_j})^2.
\end{align*}
{ Let $\theta_0^{(0)}, \theta_1^{(0)}, r^{(0)}$ be null model parameters, $u_0=(\theta_0^{(0)}, \theta_1^{(0)}, r^{(0)})^T$ and $\theta_y^{(0)} = Arg\left(\beta_0^{(0)}\frac{x+\beta_1^{(0)}}{1+\overline{\beta_1^{(0)}}x}\varepsilon\right)$, where $\beta_0^{(0)}=e^{i\theta_0^{(0)}}$ and $\beta_1^{(0)}=r^{(0)}e^{i\theta_1^{(0)}}$.}
Defining
\begin{align*}
    A_j( \theta_0, \theta_1, r) &=\frac{r^2 \cos ( \theta_0+2  \theta_1- \theta_{x_j}- \theta_y^{(0)})+2 r \cos ( \theta_0+ \theta_1- {\theta_y^{(0)}})}{r^2+2 r \cos ( \theta_1- \theta_{x_j})+1}\\
    &+\frac{\cos ( \theta_0+ \theta_{x_j}- \theta_y^{(0)})}{r^2+2 r \cos ( \theta_1- \theta_{x_j})+1},
\end{align*}
\begin{align*}
    B_j(\theta_0,\theta_1, r)&=-\frac{r^2 \sin (\theta_0+2 \theta_1-\theta_{x_j}-\theta_y^{(0)})+2 r \sin (\theta_0+\theta_1-\theta_y^{(0)})}{r^2+2 r \cos (\theta_1-\theta_{x_j})+1}\\
    &-\frac{\sin (\theta_0+\theta_{x_j}-\theta_y^{(0)})}{r^2+2 r \cos (\theta_1-\theta_{x_j})+1},
\end{align*}
we have $\theta_{\widehat{\varepsilon_j}} = \atan2(B_j(\widehat{u}), A_j(\widehat{u}))$ and $\theta_{\varepsilon_j} = \atan2(B_j({u_0}), A_j({u_0})).$
Further, a Taylor expansion yields
\begin{align}\label{atan2_taylor}
    \theta_{\widehat{\varepsilon_j}} = \theta_{\varepsilon_j} + \nabla(\atan2)(B_j(u),A_j(u))\nabla(B_j,A_j)(u) \big|_{u=u_0}(\widehat{u}-{u_0}) + r_j,
\end{align}
{where \begin{align*}
    r_j&=\frac{1}{2}(\widehat{u}-u_0)^T\nabla^2(\atan2(B_j(u),A_j(u))\big|_{u=u^*_0}(\widehat{u}-u_0)
    \\&=(\widehat{u}-u{{_0}})^TM_j(u^*_0)(\widehat{u}-u{{_0}})
    =O{{_p}}(1/n),
\end{align*}
{ $\nabla^2$ represents the Hessian matrix,} and $u^*_0$ {is between $\widehat{u}$ and $u_0$}. 
The explicit expression for $M_j(u^*_0)$ is cumbersome, so we do not show it here. However, {using the following expressions and further tedious calculations,} it can be shown that it is bounded. Expressions for $\nabla(\atan2(B_j(u),A_j(u))$ and $\nabla(B_j,A_j)(u)$ in (\ref{atan2_taylor}) are given by
}
\begin{align*}
\nabla(\atan2)(B_j(u),A_j(u))= & \begin{pmatrix}
\frac{\partial \atan2}{\partial B}(B_j(u),A_j(u))&\frac{\partial \atan2}{\partial A}(B_j(u),A_j(u))
\end{pmatrix}
\text{ and }\\
\nabla(B_j,A_j)(u)= & \begin{pmatrix}
\frac{\partial B_j}{\partial \theta_0}(u)&\frac{\partial B_j}{\partial \theta_1}(u)&\frac{\partial B_j}{\partial r}(u)\\
\frac{\partial A_j}{\partial \theta_0}(u) & \frac{\partial A_j}{\partial \theta_1}(u) & \frac{\partial A_j}{\partial r}(u),
\end{pmatrix} 
\end{align*}
where
\begin{align*}
\frac{\partial B_j}{\partial \theta_0}(u) &=-\frac{r^2 \cos (\theta_0+2 \theta_1-\theta_{x_j}-\theta_y^{(0)})+2 r \cos (\theta_0+\theta_1-\theta_y^{(0)})+\cos (\theta_0+\theta_{x_j}-\theta_y^{(0)})}{r^2+2 r \cos (\theta_1-\theta_{x_j})+1},\\
    \frac{\partial B_j}{\partial \theta_1}(u) &= -\frac{2 r^2 \cos (\theta_0+2 \theta_1-\theta_{x_j}-\theta_y^{(0)})+2 r \cos (\theta_0+\theta_1-\theta_y^{(0)})}{r^2+2 r \cos (\theta_1-\theta_{x_j})+1}\\-&\frac{2 r \sin (\theta_1-\theta_{x_j}) \left(r^2 \sin (\theta_0+2 \theta_1-\theta_{x_j}-\theta_y^{(0)})+2 r \sin (\theta_0+\theta_1-\theta_y^{(0)})\right)}{\left(r^2+2 r \cos (\theta_1-\theta_{x_j})+1\right)^2}\\
    \\-&\frac{2 r \sin (\theta_1-\theta_{x_j}) \left(\sin (\theta_0+\theta_{x_j}-\theta_y^{(0)})\right)}{\left(r^2+2 r \cos (\theta_1-\theta_{x_j})+1\right)^2}, \\
    \frac{\partial B_j}{\partial r}(u) &=  \frac{(2 r+2 \cos (\theta_1-\theta_{x_j})) \left(r^2 \sin (\theta_0+2 \theta_1-\theta_{x_j}-\theta_y^{(0)})+2 r \sin (\theta_0+\theta_1-\theta_y^{(0)})\right)}{(r^2+2 r \cos (\theta_1-\theta_{x_j})+1)^2}\\
    &+  \frac{(2 r+2 \cos (\theta_1-\theta_{x_j})) \left(\sin (\theta_0+\theta_{x_j}-\theta_y^{(0)})\right)}{(r^2+2 r \cos (\theta_1-\theta_{x_j})+1)^2}\\
    &-\frac{2 r \sin(\theta_0+2 \theta_1-\theta_{x_j}-\theta_y^{(0)})+2 \sin(\theta_0+\theta_1-\theta_y^{(0)})}{r^2+2 r \cos(\theta_1-\theta_{x_j})+1},\\
    \frac{\partial A_j}{\partial \theta_0}(u) &= \frac{r^2 (-\sin (\theta_0+2 \theta_1-\theta_{x_j}-\theta_y^{(0)}))-2 r \sin (\theta_0+\theta_1-\theta_y^{(0)})-\sin (\theta_0+\theta_{x_j}-\theta_y^{(0)})}{r^2+2 r \cos (\theta_1-\theta_{x_j})+1},\\
\frac{\partial A_j}{\partial \theta_1}(u) &= \frac{2 r \sin (\theta_1-\theta_{x_j}) \left(r^2 \cos (\theta_0+2 \theta_1-\theta_{x_j}-\theta_y^{(0)})+2 r \cos (\theta_0+\theta_1-\theta_y^{(0)})\right)}{\left(r^2+2 r \cos (\theta_1-\theta_{x_j})+1\right)^2}\\
&+ \frac{2 r \sin (\theta_1-\theta_{x_j}) \left(\cos (\theta_0+\theta_{x_j}-\theta_y^{(0)})\right)}{\left(r^2+2 r \cos (\theta_1-\theta_{x_j})+1\right)^2}\\
&+\frac{-2 r^2 \sin (\theta_0+2 \theta_1-\theta_{x_j}-\theta_y^{(0)})-2 r \sin (\theta_0+\theta_1-\theta_y^{(0)})}{r^2+2 r \cos (\theta_1-\theta_{x_j})+1}, \\
    \frac{\partial A_j}{\partial r}(u) &= \frac{2 r \cos (\theta_0+2 \theta_1-\theta_{x_j}-\theta_y^{(0)})+2 \cos (\theta_0+\theta_1-\theta_y^{(0)})}{r^2+2 r \cos (\theta_1-\theta_{x_j})+1}
    \\&-\frac{(2 r+2 \cos (\theta_1-\theta_{x_j})) \left(r^2 \cos (\theta_0+2 \theta_1-\theta_{x_j}-\theta_y^{(0)})+2 r \cos (\theta_0+\theta_1-\theta_y^{(0)})\right)}{\left(r^2+2 r \cos (\theta_1-\theta_{x_j})+1\right)^2}
    \\&-\frac{(2 r+2 \cos (\theta_1-\theta_{x_j})) \left(\cos (\theta_0+\theta_{x_j}-\theta_y^{(0)})\right)}{\left(r^2+2 r \cos (\theta_1-\theta_{x_j})+1\right)^2}.
\end{align*}
%By applying the central limit theorem and using Assumption B %for martingale differences (see e.g. McLeish (1974)) we get that $\frac{1}{\sqrt{n}} \sum\limits_{i=1}^n L_2\left(\theta_{\varepsilon_j}; u\right)$ converges weakly to a zero mean Gaussian random vector. \bk{wrong place, move}
Further, we obtain
\begin{align*}
    \Big|\frac{1}{\sqrt{n}}\sum_{j=1}^n & R_{1j}(t)\Big| 
    \leq \frac{1}{2n^{\frac{3}{2}}}t^2\sum_{j=1}^n \Big|\sqrt{n}(\theta_{\widehat{\varepsilon_j}}-\theta_{\varepsilon_j})\Big|^2 \\
    &=  \frac{1}{2n^{\frac{3}{2}}}\sum_{j=1}^n t^2\Big|\Big(\nabla(\atan2)(B_j(u),A_j(u))\nabla(B_j,A_j)(u) \Big)\Big|_{u=u_0} \sqrt{n} (\widehat{u}-u_0) + \sqrt{n}r_j\Big|^2\\ 
    &\leq \frac{1}{2n^{\frac{3}{2}}}\sum_{j=1}^n t^2||\Big(\nabla(\atan2)(B_j(u),A_j(u))\nabla(B_j,A_j)(u) \Big)\Big|_{u=u_0} ||^2||\sqrt{n} (\widehat{u}-u_0)||^2\\& + \frac{1}{2\sqrt{n}}\sum_{j=1}^n t^2 ||r_j||^2=t^2O_p\Big(\frac{1}{\sqrt{n}}\Big),
\end{align*}
and, similarly, $\Big|\frac{1}{\sqrt{n}}\sum\limits_{j=1}^n R_{2j}(t)\Big| \leq t^2O_p\left(1/\sqrt{n}\right)$.
Applying the delta method yields 
\begin{align*}
    \sqrt{n}(\hat{\delta^t}-\delta^t)=&\sqrt{n}t\delta^{t-1}(\hat{\delta}-\delta)+o_p(1)
    =t\delta^{t-1}(\frac{1}{\sqrt{n}} \sum_{i=1}^n L_1(\theta_{\varepsilon_j};\delta)+r_\delta)+o_p(1).
\end{align*}
By assumptions A and C, $\|t\delta^{t-1}r_\delta\|_w=o_p(1).$
%\bk{Since $\delta\in[0,1]$ and $\|t\|_w<\infty$, this should be obvious (if $||\cdot||$ is the same as $\|\cdot\|_w$)
%
%\bk{The weight function is $\omega$, but in section 3, you use $w$, this should be fixed}
Combining the results, we obtain
\begin{align*}
    &I_n^{(1)}(t)+I_n^{(2)}(t)
    =\frac{1}{\sqrt{n}} \sum_{j=1}^{n} \Big(\cos (t \theta_{\varepsilon_j})+i\sin (t \theta_{\varepsilon_j})-\delta^t\\&\quad+t (\theta_{\widehat{\varepsilon_j}}-\theta_{\varepsilon_j})\left(i\cos(t\theta_{\varepsilon_j})-\sin(t\theta_{\varepsilon_j})\right) +t\delta^{t-1}L_1(\theta_{\varepsilon_j};\delta)\Big) + o_p(1)\\
    \quad
    %&= \frac{1}{\sqrt{n}} \sum_{j=1}^{n} \Big(\cos (t \theta_{\varepsilon_j})+i\sin (t \theta_{\varepsilon_j})-\delta^t\\
    %&+\frac{t}{\sqrt{n}} \Big(\nabla(\atan2)(B_j(u),A_j(u))\nabla(B_j,A_j)(u) \Big)\Big|_{{u=u_0}}\\&\cdot \sqrt{n}(\widehat{u}-u_0)(i\cos(t\theta_{\varepsilon_j})-\sin(t\theta_{\varepsilon_j})) +t\delta^{t-1}L_1(\theta_{\varepsilon_j};\delta)\Big) + o_p(1)\\ 
    =& \frac{1}{\sqrt{n}} \sum_{j=1}^{n} \Big(\cos (t \theta_{\varepsilon_j}) +i\sin (t \theta_{\varepsilon_j}) -\delta^t\\
    &\quad+ it \Big(\nabla(\atan2)(B_j(u),A_j(u))\nabla(B_j,A_j)(u) \Big)\Big|_{{u=u_0}} \frac{1}{n}\sum\limits_{k=1}^n L_2\left(\theta_{\varepsilon_k} ; u_0\right)\cos(t\theta_{\varepsilon_j})\\&\quad-t \Big(\nabla(\atan2)(B_j(u),A_j(u))\nabla(B_j,A_j)(u) \Big)\Big|_{u=u_0} \frac{1}{n}\sum\limits_{k=1}^n L_2\left(\theta_{\varepsilon_k} ; u_0\right)\sin(t\theta_{\varepsilon_j})\\&\quad+t\delta^{t-1}L_1(\theta_{\varepsilon_j};\delta)\Big) + r_{tot} (t),
    \end{align*}
    %\bk{index $j$ in $r_{tot,j}$ removed}
    {where $r_{tot}(t)$ is the sum of the remainders that appear in the approximations used so far, satisfying} 
    \begin{align*}
       | \frac{1}{\sqrt{n}}\sum_{j=1}^nr_{tot,j}(t)|\leq t^2o_p(1).
    \end{align*}
Let $C(\theta_{\varepsilon_j}) = \Big(\nabla(\atan2)(B_j(u),A_j(u))\nabla(B_j,A_j)(u) \Big)\Big|_{u=u_0}$ and $$\Phi^{(1)}(\theta_{\varepsilon_j},\theta_{\varepsilon_k}{;t}) = \frac{1}{2}(C(\theta_{\varepsilon_k})L_2(\theta_{\varepsilon_j};u_0)\cos(t\theta_{\varepsilon_k})
 + C(\theta_{\varepsilon_j}) L_2(\theta_{\varepsilon_k};u_0)\cos(t\theta_{\varepsilon_j})),$$
$$\Phi^{(2)}(\theta_{\varepsilon_j},\theta_{\varepsilon_k}{;t}) = \frac{1}{2}(C(\theta_{\varepsilon_k})L_2(\theta_{\varepsilon_j};u_0)\sin(t\theta_{\varepsilon_k})
 + C(\theta_{\varepsilon_j}) L_2(\theta_{\varepsilon_k};u_0)\sin(t\theta_{\varepsilon_j})).$$
Then, if we denote $V_n^{(1)}(t) = \frac{1}{n^2} \sum\limits_{j=1}^n\sum\limits_{k=1}^n \Phi^{(1)}(\theta_{\varepsilon_j},\theta_{\varepsilon_k}{;t})$ and $V_n^{(2)}({t}) = \frac{1}{n^2} \sum\limits_{j=1}^n\sum\limits_{k=1}^n \Phi^{(2)}(\theta_{\varepsilon_j},\theta_{\varepsilon_k}{;t})$, using Hoeff\-ding's theorem for the asymptotic distribution of V-statistics, we get that $\sqrt{n}V_n^{(i)}({t}) \stackrel{{\cal L}}{\to} \mathcal{N}(0, 4\sigma_{(i)}^2({t}))$, where $\sigma_{(i)}^2({t}) = E(\Phi_1^{(i)}(\theta_{\varepsilon};{t}))^2, i=1,2$, and 
\begin{align*}
\Phi_1^{(1)}(x;{t}) &= E\Phi^{(1)}(x,\theta_{\varepsilon}
;{t})=\frac{1}{2}E\Big(C(\theta_{\varepsilon})L_2(x;u_0)\cos(t\theta_{\varepsilon})\Big), \\
\Phi_1^{(2)}(x;{t}) &=E\Phi^{(2)}(x,\theta_{\varepsilon};{t} )=\frac{1}{2}E\Big(C(\theta_{\varepsilon})L_2(x;u_0)\sin(t\theta_{\varepsilon})\Big)
\end{align*}
are the first projections of kernels $\Phi^{(1)}$ and $\Phi^{(2)}$, respectively. { It  can be numerically shown, taking into account that $L_2(\cdot,u_0)\neq \mathbf{0}$, that these projections are nonconstant functions. } 
Then, we have 
\begin{align*}
    I_n^{(1)}(t)+  I_n^{(2)}  (t) 
    &=
    \frac{1}{\sqrt{n}} \sum_{j=1}^{n} \Big(\cos (t \theta_{\varepsilon_j})+i\sin (t \theta_{\varepsilon_j})-\delta^t+t\delta^{t-1}L_1(\theta_{\varepsilon_j};\delta)\Big)\\&\;\;\;+ it 
    \sqrt{n} V_n^{(1)}({t})-t\sqrt{n} V_n^{(2)}({t})+ {r_{tot}(t)}
    \\ 
   % &=\frac{1}{\sqrt{n}} \sum_{j=1}^{n} \Big(\cos (t \theta_{\varepsilon_j})+i\sin (t \theta_{\varepsilon_j})-\delta^t+t\delta^{t-1}L_1(\theta_{\varepsilon_j};\delta)\Big)\\&\;\;\; + it 
    %\frac{2}{\sqrt{n}}\sum_{j=1}^n \Phi_1^{(1)}(\theta_{\varepsilon_j};{t})  -t\frac{2}{\sqrt{n}}\sum_{j=1}^n \Phi_1^{(2)}(\theta_{\varepsilon_j};{t}) + {R'(t)} \\
    &= \frac{1}{\sqrt{n}} \sum_{j=1}^{n} \Big(\cos (t \theta_{\varepsilon_j})+i\sin (t \theta_{\varepsilon_j})-\delta^t+t\delta^{t-1}L_1(\theta_{\varepsilon_j};\delta)\\&\;\;\;+ 2it 
    \Phi_1^{(1)}(\theta_{\varepsilon_j};{t})-2t \Phi_1^{(2)}(\theta_{\varepsilon_j};{t})\Big) + {R'(t)}    \\
    &= \frac{1}{\sqrt{n}} \sum_{j=1}^{n} \Gamma (t,\theta_{\varepsilon_j}; u_0, \delta){ + {R'(t)}}.
    \end{align*}
From the bound for $r_{tot}(t)$ and the Hoeffding approximation of the V-statistic, it follows that $|R'(t)|\leq o_p(1)t^2$.
Furthermore, due to the fact that $L_1$ and $L_2$ have expectation 0, it can be seen that $E\{\Gamma(t,\theta_{\varepsilon_1};u_0,\delta)\}=0.$
Finally, applying the central limit theorem in Hilbert spaces (see e.g. \cite{bosq2000linear}), we obtain that the limiting process is a centered Gaussian with covariance kernel
\begin{align}\label{covkernel}
K_{\mathcal{G}}(s, t)  = \mathbb{E}\{\Gamma(t,\theta_{\varepsilon};u_0,\delta)\overline{\Gamma}(s,\theta_{\varepsilon};u_0,\delta)\},
\end{align}
    where %\begin{align*}
    $\Gamma(t,\theta_{\varepsilon};u_0,\delta)=\cos (t \theta_{\varepsilon})+i\sin (t \theta_{\varepsilon})-\delta^t+t\delta^{t-1}L_1(\theta_{\varepsilon};\delta)+ 2it 
    \Phi_1^{(1)}(\theta_{\varepsilon};t)-2t \Phi_1^{(2)}(\theta_{\varepsilon};t).$%\end{align*}
 The statement of Theorem 1 is now a direct consequence of the continuous mapping theorem.
\end{proof}

\section{Circular densities of the alternative distributions} \label{alternatives}
\begin{itemize}
 \item wrapped normal distribution $\mathcal{WN}(\mu,\rho)$
    $$ f(\theta)=\frac{1}{2\pi}\Big(1+2\sum_{k=1}^{\infty}\rho^{k^2}\cos k(\theta-\mu)\Big), \; 0\leq \theta,\mu <2\pi,\; 0<\rho<1.$$
    \item von Mises distribution $\mathcal{VM}(\mu,k)$
    $$f(\theta)=\frac{1}{2\pi I_0(k)}e^{k\cos(\theta-\mu)}\textup,\;  0\leq \theta,\mu < 2\pi, \; k>0,$$ where $I_0(k)$ is the modified Bessel function of the first kind.

     \item cardioid distribution $\mathcal{C}a(\mu,\rho)$
    $$f(\theta) = \frac{1}{2\pi}(1 + 2\rho \cos(\theta - \mu)), \; 0\leq \theta, \mu <2\pi, |\rho|<\frac{1}{2},$$
    \item Cartwright's power-of-cosine distribution $\mathcal{CW}$($\mu, \zeta$)
    $$f(\theta) = \frac{2^{\frac{1}{\zeta}-1}\Gamma^2\left(\frac{1}{{\zeta}} + 1\right)(1 + \cos(\theta - \mu))^{1/{\zeta}}}{\pi\Gamma\left(\frac{2}{\zeta} + 1\right)}, \; 0 \leq \theta,\mu < 2\pi, \; \zeta>0,$$
    where $\Gamma(\cdot)$ is the gamma function.
    \item Jones-Pewsey distribution $\mathcal{JP}(\mu,\kappa,\psi$)
    $$f(\theta) = \frac{\cosh(\kappa\psi) + \sinh(\kappa\psi) \cos(\theta - \mu)^{\frac{1}{\psi}}}{2\pi P_{\frac{1}{\psi}}(\cosh(\kappa\psi))}, \; 0 \leq \theta,\mu < 2\pi, \; \kappa\geq 0, \; \psi \in \mathbb{R},$$
    where \(P_{\frac{1}{\psi}}(\cdot)\) is the associated Legendre function of the first kind with degree \(\frac{1}{\psi}\) and order 0.
    \item Batschelet distribution $\mathcal{B}a(\mu,\kappa,\nu$)
    $$f(\theta)=\frac{1}{B_0(\kappa,\nu)}e^{\kappa \cos((\theta-\mu)+\nu \sin(\theta - \mu))}, \; 0 \leq \theta,\mu < 2\pi, \; \kappa\geq 0, \;-1 \leq \nu \leq 1,$$
    where $$B_p(\kappa,\nu)=\int_0^{2\pi} \cos(p\theta)e^{\kappa \cos(\theta+\nu \sin\theta)}d\theta, \; p = 0, 1, ...$$
  
\end{itemize}

\section{Classical bootstrap algorithm} \label{app:class-boot}
\begin{enumerate}
     \item For fixed $x_1, x_2,\ldots, x_n$ and $y_1, y_2,\ldots, y_n$ compute the estimator $\widehat{v}_n:=\widehat{v}_n(x_1, x_2,\ldots, x_n;y_1, y_2,\ldots, y_n)$ of the parameter vector $v=(\theta_0,\theta_1,r,\delta)$.
    \item Calculate the residuals $\widehat{\theta}_{\varepsilon}$ of the proposed regression model, using $\hat{y} = \widehat{\beta}_0 \frac{x+\widehat{\beta}_1}{1+\overline{\widehat{\beta}_1}x}$.
    \item Compute the test statistic of interest, say $S_n=S_n(\widehat{\theta}_{\varepsilon_1}, \widehat{\theta}_{\varepsilon_2},\ldots, \widehat{\theta}_{\varepsilon_n};\widehat{v}_n)$.
    \item Repeat the following steps for each $i\in \{1, ..., B\}$:
    \begin{enumerate} 
        \item[a)] Generate a bootstrap sample $y_1^*, y_2^*,\ldots, y_n^*$, where $y_j^*=\widehat{\beta}_0 \frac{x+\widehat{\beta}_1}{1+\overline{\widehat{\beta}_1}x}\varepsilon_j^*$ and $\varepsilon_j^*\sim WC(\widehat{\delta})$.
        \item[b)] On the basis of $(y_1^*,\ldots,y_n^*)$, obtain a bootstrap estimate $\widehat{v}_n^*$ of $v$.
        \item[c)] Calculate the residuals $\widehat{\theta}_{\varepsilon}^*$ of the proposed regression model, using $\hat{y}^* = \widehat{\beta}_0^* \frac{x+\widehat{\beta}_1^*}{1+\overline{\widehat{\beta}_1^*}x}$.
        \item[d)] Compute the value of the bootstrap test statistic $S_n^{*(i)}=S_n(\widehat{\theta}_{\varepsilon_1}^*, \widehat{\theta}_{\varepsilon_2}^*,\ldots, \widehat{\theta}_{\varepsilon_n}^*;\widehat{v}_n^*)$. 
    \end{enumerate}
    \item Reject the null hypothesis if $S_n\geq c_\alpha$, where $c_\alpha$ denotes the $(1-\alpha)\%$ quantile of the empirical distribution of the bootstrap test statistics $(S_n^{*(1)},\ldots,S_n^{*(B)})$. 
\end{enumerate}

\section{Warp-speed bootstrap algorithm}
\begin{enumerate}
     \item For fixed covariates $x_1, x_2,\ldots, x_n$, simulate a sequence $\varepsilon_1, \varepsilon_2,\ldots,\varepsilon_n\sim F_1$, where $F_1$ denotes one of the alternative distributions defined above, and the corresponding $y_1, y_2,\ldots, y_n$.  
    \item Compute the estimator $\widehat{v}_n:=\widehat{v}_n(x_1, x_2,\ldots, x_n;y_1, y_2,\ldots, y_n)$ of the parameter vector $v=(\theta_0,\theta_1,r,\delta)$.
    \item Calculate the residuals $\widehat{\theta}_{\varepsilon}$ of the proposed regression model, using $y_{fit} = \widehat{\beta}_0 \frac{x+\widehat{\beta}_1}{1+\overline{\widehat{\beta}_1}x}$.
    \item Compute the test statistic of interest, say $S_n=S_n(\widehat{\theta}_{\varepsilon_1}, \widehat{\theta}_{\varepsilon_2},\ldots, \widehat{\theta}_{\varepsilon_n};\widehat{v}_n)$.
    \item Generate a bootstrap sample $y_1^*, y_2^*,\ldots, y_n^*$, where $y_j^*=\widehat{\beta}_0 \frac{x+\widehat{\beta}_1}{1+\overline{\widehat{\beta}_1}x}\varepsilon_j^*$ and $\varepsilon_j^*\sim WC(\widehat{\delta})$.
    \item On the basis of $(y_1^*,\ldots,y_n^*)$, obtain a bootstrap estimate $\widehat{v}_n^*$ of $v$.
    \item Calculate the residuals $\widehat{\theta}_{\varepsilon}^*$ of the proposed regression model, using $\hat{y}^* = \widehat{\beta}_0^* \frac{x+\widehat{\beta}_1^*}{1+\overline{\widehat{\beta}_1^*}x}$.
    \item Compute the value of the bootstrap test statistic $S_n^*=S_n(\widehat{\theta}_{\varepsilon_1}^*, \widehat{\theta}_{\varepsilon_2}^*,\ldots, \widehat{\theta}_{\varepsilon_n}^*;\widehat{v}_n^*)$.
    \item Repeat the previous steps $B$ times and thereby produce two sequences of test statistics: $\{S_n^{(1)}, S_n^{(2)},\ldots,S_n^{(B)}\}$ and $\{S_n^{*(1)}, S_n^{*(2)},\ldots,S_n^{*(B)}\}$.
    \item Reject the null hypothesis for the $j$-th sample $(j=1,\ldots,B),$ if $S_n^{(j)}\geq c_\alpha$, where $c_\alpha$ denotes the $(1-\alpha)\%$ quantile of the empirical distribution of the bootstrap test statistics $(S_n^{*(1)},\ldots,S_n^{*(B)})$. 
\end{enumerate}
\end{appendix}
\bibliographystyle{abbrv}
\bibliography{ARXIV}

\end{document}